\documentstyle[aps,preprint]{revtex}
\begin{document}
\draft
\preprint{IMSc-95/25;~ hep-th/9512209}
\title{Eikonal Particle Scattering and Dilaton Gravity}
\author{Saurya Das and Parthasarathi Majumdar
\footnote{E-Mail:~saurya,partha@imsc.ernet.in}}
\address{The Institute of Mathematical Sciences, \\ CIT Campus,
Madras - 600 113,  India.}
\maketitle
\begin{abstract}
Approximating light charged point-like particles in terms of
(nonextremal) dilatonic black holes is shown to lead to certain pathologies in
Planckian scattering in the
eikonal approximation, which are traced to the presence of a (naked) curvature
singularity in the metric of these black holes. The existence of such
pathologies is
confirmed
by analyzing the problem in an `external metric' formulation where an
ultrarelativistic
point particle scatters off a dilatonic black hole geometry at large impact
parameters. The maladies disappear almost trivially upon imposing the extremal
limit.
Attempts to
derive an effective three dimensional `boundary' field theory
in the eikonal limit are stymied by four dimensional (bulk)
terms proportional to the light-cone derivatives of the dilaton field, leading
to
nontrivial mixing of electromagnetic and gravitational effects, in contrast to
the case
of general relativity. An eikonal scattering amplitude, showing decoupling of
these
effects, is shown to be derivable by resummation of graviton, dilaton and
photon exchange
ladder diagrams in a linearized version of the theory, for an asymptotic value
of the
dilaton field which makes the string coupling constant non-perturbative.
\end{abstract}
%\pacs{04.60. -m, 04.62. +v, 11.80.Fv}
\newpage

\section{Introduction}

Nontrivial nonperturbative information regarding gravitational interactions is
now
well-known to be accessible via point particle scattering in four dimensional
Minkowski
space at
Planckian centre-of-mass energies and fixed, low momentum transfers
\cite{thf,acv}.
The
singular kinematics of this (eikonal) approximation lead to a truncated
dynamics amenable to
exact treatment without further approximations. The easiest way to visualize
these
collision processes is through the {\it shock wave} picture \cite{thf,aich},
wherein an
ultrarelativistic point particle produces a background which has the geometry
of two
Minkowski spacetimes glued together after a shift along the null direction
(in Minkowski space) characterizing the motion of the particle \cite{thdr}. The
other null direction
can be taken to define the affine parameter for the null geodesic of a test
particle
encountering this shock wave geometry. The quantum mechanical amplitude of this
collision
is exactly calculable, so long as $Gs~\approx~1$, where $G$ is Newton's
constant. A field
theoretic analysis reproduces identical results for the amplitudes while
yielding a
reduced {\it three} dimensional field theory which describes the suppression of
standard graviton exchanges relative to the instantaneous interaction mediated
by the shock
wave \cite{ver}. Leading order corrections to the eikonal process have also
been computed
using superstring theory in the Regge-Gribov formalism \cite{acv}.

The
inclusion of and interplay (of gravitation) with electromagnetism, in this
kinematical domain, has also been investigated in detail
\cite{jac,dm1,dm2,dm3,dm4},
incorporating situations where the particles may have both electric and
magnetic charge.
In so far as general relativity is concerned, some remarkable phenomena occur
in the
eikonal region: electromagnetic and gravitational interactions seem to operate
quite
independent of each other, in contrast to more generic kinematical situations
\cite{dm2}.
\footnote{This has been further confirmed in an independent analysis using the
external
metric
formulation of the problem, wherein an almost luminal particle scatters off the
static
metric of a charged (Reissner-Nordstr\"om) black hole \cite{dm4}.} Also, while
gravitational interactions characterized by the dimensionless quantity $Gs$
is usually taken to dominate in this region ($Gs \approx 1$), compared to
electromagnetism
which is
controlled by $\alpha \approx 1/137$ for small momentum transfers, with
magnetic charges
present this is no longer the case \cite{dm1}.

The variant of general relativity known as dilaton gravity is an important
extension of
the standard Einstein theory because it appears in the low energy approximation
to
superstring theory \cite{gsw}. The behavior of dilaton gravity in the
kinematics of the
eikonal
approximation is a question of intrinsic interest vis-a-vis the simplifications
mentioned
above. On somewhat heuristic grounds, it has been shown \cite{dm2} that the
decoupling of
gravity and electromagnetism seen earlier may not actually occur for the case
of dilaton
gravity, owing primarily to
the coupling of the dilaton field to the metric (or to the electromagnetic
field
strength). In this paper we turn to a more comprehensive analysis of dilaton
gravity in
the eikonal domain, to see if these heuristic results may indeed have a firmer
basis.
Thus, if the particles in question are approximated in their static limit by
charged dilatonic black holes, then is the geometry due to such a particle
similar to a
gravitational shock wave when the particle moves almost luminally? The issue of
the
eikonal scattering amplitude in this case is an immediate consequence. The
reduction of
the full set of degrees of freedom to a truncated set amenable to exact
mathematical
treatment is another issue of importance that must be addressed.

The paper is organized as follows: in section \ref{heurist} we review our
earlier work using the
boosting techniques of ref. \cite{thdr} to examine the interplay of gravity and
electromagnetism. We further demonstrate how the problems discerned might
disappear in the
extremal limit. In section \ref{external}, both the non-extremal and the
extremal situations are
re-analyzed within the external (static dilatonic
black hole) metric formalism; in the former case, we show how it is impossible
to reduce
the equation
of motion of an ultrarelativistic particle in this metric to a solvable
Schr\"odinger-like
form useful for extracting phase shifts. Once again, the pathology is obviated
in the
extremal limit wherein solutions identical to those in a Schwarzschild
background
\cite{dm4} ensue. In section \ref{sca}, we turn to a field theoretic
analysis following \cite{ver}, wherein we point out the difficulties of
reducing the
theory in the relevant kinematical domain to a boundary field theory which
`lives' in a
three dimensional
space composed by the transverse two dimensional plane and the boundary of the
null plane.
This concomitantly demonstrates the nontrivial mixing of gravitational and
electromagnetic
interactions in this case. Section \ref{pert} probes the possibility of a
derivation of
the quantum eikonal amplitude by resummation of ladder-type exchange graphs in
a
linearized version of the theory. The linearization is argued to be invalid in
the
regime of perturbative string coupling. We conclude in section \ref{concl}
with a few remarks on
what our results might indicate from a string theoretic standpoint.

\section{Dilaton Gravity Heuristics}
\label{heurist}

This section is a brief review of our earlier work \cite{dm2}. We begin by
considering the static, spherically symmetric and electrically charged
solution of dilaton gravity in the so-called `string metric' \cite{gar},
which is a solution of the low energy string effective action:
\begin{equation} ds^2~=~ (1-{\alpha \over Mr} )^{-1} \left [ (1 - {2GM
\over r}) dt^2~-(1 -{2GM \over r})^{-1} dr^2~-~(1 -{\alpha \over Mr}) r^2
d \Omega^2 \right ]~~. \label{metr} \end{equation}
Here $\alpha \equiv Q^2 e^{2\phi_0}$,
$Q$ being the
electric charge and $\phi_0$ the asymptotic value of the dilaton field.
We confine ourselves
to situations {\it not} subject to the extremality condition $Q^2
e^{2\phi_0} = 2GM^2$. It may be noted that this
metric differs from the Reissner-Nordstr\"om solution of
general relativity in that it does not have two horizons, while it has a
curvature singularity at $r=\alpha/M$.
This difference is due to the
presence of the dilaton field.
As this metric describes the spacetime around a point particle of
mass $M$, to obtain the same when the particle is massless and
travels along the null geodesic $x^- \equiv t-z =0$,
we boost this metric
along the positive $z$ axis to a velocity $\beta$ and take the
limit $\beta \rightarrow 1$.
On parametrizing the mass as $M=p/\gamma$, where $\gamma
=(1-\beta^2)^{-1/2}$ and $p$ is the energy of the particle, and
introducing the other light cone coordinate $x^+=t+z$, we
get \cite{dm2}:
$$ds^2~\rightarrow~d{\tilde x}^+ d{\tilde x}^-~-~(d{\tilde
x}_{\perp})^2~~, $$ where,
\begin{eqnarray}
d{\tilde x}^+ ~~&=&~~dx^+ ~-~\left ( { {4Gp \over |x^-|} \over {1 - {\alpha
\over p |x^-|}}} \right ) dx^-  \nonumber \\
d{\tilde x}^-~~&=&~~dx^- \left ( { {1 - {\alpha \over 2p |x^-|}} \over {1 -
{\alpha \over p |x^-|}} } \right )  \nonumber \\
d{\tilde {\vec x}}_{\perp}~~&=&~~d{\vec x}_{\perp}~~. \label{shft}
\end{eqnarray}
We observe that in addition to the shift in the $x^+$ coordinate
(as for the Schwarzschild metric),
the coordinate $\tilde x^-$, depends on the charge
$\alpha$. This is made explicit by choosing $\alpha$ to be small
(achieved either by considering a small charge $Q$ or by taking
a large negative value of $\phi_0$). Then the above equations can
be linearized to obtain,
\begin{eqnarray}
d{\tilde x}^+~~&=&~~dx^+ ~-~{4Gp \over |x^-|} ~-~{4\alpha \over (x^-)^2}
{}~+~ {\cal O}(\alpha ^2 / p)~ \label{xpls} \\
d{\tilde x}^-~~&=&~~dx^-~+~{\alpha \over 2p |x^-|} ~+~ {\cal O}( \alpha^2 /
p^2)~. \label{xmns}
\end{eqnarray}
The $\alpha$ dependent
shift in $x^+$, being a continuous function of $x^-$,
can be removed by a diffeomorphism while the shift in $x^-$
cannot, because of the presence of the discontinuous function
$\theta(x^-)$. Interestingly, for the Reissner-Nordstr\"om metric,
the $Q$ dependent piece can also be removed by a diffeomorphism.
Now, for a test particle in the background geometry of this right-moving
particle, the coordinate $x^-$ serves as its affine parameter, and
a discontinuity in the latter signals a serious breakdown of the
boosting method. Specifically, the interpretation of the
boosted metric as two Minkowski spaces glued together
at the null plane $x^-=0$ after a shift in the coordinate $x^+$
(cut and paste prescription)
is
no longer possible as for the Schwarzschild \cite{thdr} or the
Reissner-Nordstr\"om metric \cite{dm2} metric.
This becomes apparent when one writes the
classical geodesic equations for a light test particle in the
background of the boosted dilaton metric and tries to solve it
perturbatively in a power series in the mass $M$ using singular
perturbation theory. The failure of
the latter indicates that the the null geodesics are incomplete in
this case and curvature singularity at
$r=\alpha/M$ shows up as an extended
naked singularity in the boosted limit \cite{dm2}. Thus
the geometry is intractably more complicated which renders a
calculation of the corresponding scattering amplitude impossible.

Having confronted the above mentioned difficulty, let us try to
see whether the same can be circumvented for certain special values
of the parameters. For example, the extremal limit can be
considered for its special role in certain other situations (it
has zero entropy and Hawking temperature). For the space-time
depicted my metric (\ref{metr}), the extremal limit corresponds to
the merging of the Schwarzschild horizon and the sphere of
curvature singularity. The condition among the parameters is
therefore $\alpha = 2GM^2$, which when translated in the
expression for the metric yields,
\begin{equation}
ds^2~=~-dt^2 + \frac{dr^2}{\left(1 - \frac{2GM}{r} \right)^2} +
r^2 d\Omega^2~.
\label{extr}
\end{equation}
On performing the boosting procedure on this, we get:
\begin{equation}
ds^2=~dx_{\perp}^2 - dx^- \left[ dx^+ - 4Gp\frac{dx^-}{|x^-|}
\right]~,
\end{equation}
which can be seen to coincide with (\ref{xmns}) for $\alpha = 0$.
Note that this is the same as a boosted Schwarzschild geometry
\cite{thdr}, although the metric (\ref{extr}) cannot be
identified with a Schwarzschild space-time. In fact, this metric
is singularity free and geodesically complete.
Since there is a shift in the light cone coordinate $x^+$ only,
the affine parameter $x^-$ is continuous, and
the `cut-and-paste' prescription is eminently applicable.
The corresponding scattering amplitude is the well known
eikonal result \cite{thf}:
\begin{equation}
f(s,t)~=~{1 \over t}~\frac{\Gamma(1-iGs)}
{\Gamma(1+iGs)}\left(\frac{1}{-t}\right)^{-iGs}~,
\label{amp}
\end{equation}
where $s$ is the square of the center-of-mass energy.
It may be noted that the above amplitude refers to gravitational
interactions only. In addition, due to charges on the particles,
there can be electromagnetic contributions to the scattering. How
they affect the latter has been dealt with at length in \cite{dm2}
and \cite{dm4}. We will briefly touch upon this issue in section
\ref{sca}.
We will also come back to the issue of taking the extremal limit in the
subsequent sections and try to understand why it leads to a
reasonable result.

\section{External Metric Approach}
\label{external}

A better physical insight into why such a breakdown occurs for the
generic dilaton gravity metric may emerge upon analyzing
the above physical situation by a
manifestly covariant approach, in which we solve for the wave
equation of a test particle in the fixed background space-time
created by the other particle.
As emphasized earlier, this spacetime
can be modelled by a the dilaton black hole
solution as in Eq.(\ref{metr}). For simplicity, we define the following
quantities :
\begin{eqnarray}
\Lambda &=& 1 - \frac{2GM}{r} \nonumber \\
and~~~~\Delta &=& 1 - \frac{\alpha}{Mr}~~. \nonumber
\end{eqnarray}
The Klein-Gordon equation of the (spinless)
test particle is given by:
\begin{equation}
D_\mu D^\mu \phi~=~0~~, \label{kg}
\end{equation}
where $D_\mu$ denotes the relativistically covariant derivative in the
metric (\ref{metr}). Assuming a solution for $\phi$ of the form
\begin{equation}
\phi (\vec r,t)~=~\phi(r)~Y_{lm}(\theta, \phi)~e^{iEt}~,
\end{equation}
(where $E$ is the energy of the test particle as measured by an
asymptotic observer) and with the
`string'
metric (\ref{metr}) in the background, the radial part of (\ref{kg})
becomes :
\begin{equation}
r^2 \Lambda \frac{d^2\phi(r)}{dr^2} +
\frac{d(r^2\Lambda)}{dr}
{}~\frac{d\phi(r)}{dr} -
\left[\frac{l(l+1)}{\Delta} - \frac{E^2r^2}{\Lambda}
\right]~\phi(r) ~=~0~~\label{rad}.
\end{equation}
For generic values of $\Lambda$, the first derivative term can be
ignored and
on setting $\Delta=1$ (i.e. no dilatonic
and/or electric charge), we recover the radial equation of a
neutral particle in a Schwarzschild background \cite{dm2}:
\begin{equation}
\frac{d^2f}{dr^2} - \left[ \frac{l(l+1)-3(Gs)^2}{r^2}
-\frac{2GsE}{r} -E^2 \right]f ~=~0~.
\label{scrad}
\end{equation}
Here, $\phi(r) = f(r)/r$.
For large $l$ (the eikonal limit), this
equation is just the Sch\"odinger equation for a charge in
a Coulomb potential, once we identify the electromagnetic coupling
constant $\alpha$ with $\alpha_G \equiv Gs$ (with a minus sign)
and the momentum
$k$ with the energy $E$. The subsequent
calculation of the scattering phase shifts is exact.
The expression for the phase shift is \cite{davy,dm4}:
\begin{equation}
\delta_l~=~\arg \Gamma(l+1-iGs)~.
\label{phase}
\end{equation}
The scattering amplitude obtained form this phase shift agrees
with (\ref{amp}).
However, we are interested to know whether for generic values of
$\Delta$, the above equation
reduces to a Schr\"odinger-like equation, amenable to
scattering solutions. In the latter case,
$\Delta$ vanishes and the centrifugal term becomes singular
at a radius $r=\alpha/M$. In the limit that $M$ is small, this corresponds
to very large radial distances. Thus the curvature singularity
appears in the vicinity of the test particle trajectory (with
fixed large impact parameter $b$) and the tacit assumption that
the test particle trajectory is in a region of small curvature,
fails. This warrants a careful
analysis of the radial equation in this region.
The coefficient of $\phi(r)$ in (\ref{rad})
is \begin{equation}
p^{(2)}~\equiv~\frac{E^2r^2}{\Lambda} - \frac{l(l+1)}{\Delta}~.
\end{equation}
In the domain of interest $0 < r < \infty$, $p^{(2)}$
fails to be continuous at $r=\alpha/M$. This is because
\begin{eqnarray}
\lim_{r \rightarrow (\alpha/M)^{-}} p^{(2)} \rightarrow +\infty~~,
\nonumber \\
\lim_{r \rightarrow (\alpha/M)^{+}} p^{(2)} \rightarrow -\infty~~,
\end{eqnarray}
and $p^{(2)}|_ {r = \alpha/M}$ is not defined.
An elementary theorem in the theory of ordinary differential
equations states that, under these circumstances, a unique solution of
(\ref{rad}) does not exist \cite{ince}. Similar conclusions follow by
considering the
radial equation in the `Einstein' metric, which is related to the
string metric by a Weyl transformation of the form~~$g_{\mu
\nu}^{Einstein}=e^{2\phi}g_{\mu \nu}^{string}$. This can be seen by
writing the radial equation in
this case, which is :
\begin{equation}
r^2 \Lambda \frac{d^2\phi(r)}{dr^2} +
\left[ \frac{d(r^2\Lambda)}{dr} + \frac{r^2
\Lambda}{\Delta}~\frac{d\Delta}{dr}\right]~\frac{d\phi(r)}{dr} -
\left[\frac{l(l+1)}{\Delta} - \frac{E^2r^2}{\Lambda}
\right]~\phi(r) ~=~0~~\label{rad1}.
\end{equation}
Here, in addition to $p^{(2)}$, the coefficient $p^{(1)}$ of the
first derivative term also becomes discontinuos at $r=\alpha/M$
due to the presence of the additional $\Delta$ - dependent piece.
So, we can no longer ignore the first derivative term. In any case, a
unique solution still does not exist.

Thus we see that, for vanishing particle masses, it is
impossible to
extract a Schr\"odinger-like differential equation for the dilaton
gravity metric from which we can compute a unique scattering solution
and the corresponding phase
shift. Basically, the reason is that the factor in the metric incorporating
dilaton
effects, namely $\left(1 -
\alpha/Mr \right)$, blows up as $M
\rightarrow 0$ thus rendering the equation analytically
intractable. As the particle masses decrease,
the location of the curvature singularity
of the black hole recedes away from the origin $r=0$ further without limit.
Any particle in the field of this black hole, however large its impact
parameter,
is trapped within this naked singularity. This is
reflected in the non-existence of well-defined quantum scattering
solutions. The gulf of difference between the earlier analyses
involving the Schwarzschild and Reissner-Nordstr\"om metrics
\cite{dm4} and the present case, need hardly be over-emphasized.
The problem is obviously absent for macroscopic stellar objects with large
masses, for which the naked singularity is well hidden behind the event
horizon.
One can then expand the
coefficients of the radial equation involving $\Delta$ in powers of the small
parameter $\alpha/Mr$ and obtain a perturbative solution. This would yield {\it
finite}
$\alpha$-dependent corrections to the scattering amplitude
(\ref{amp}) which, however, detracts from our aim of studying point particle
scattering.

Instead, it makes more sense to investigate the extremal limit which was seen
to cure the malady in the
previous section. Substituting $\Lambda = \Delta$, for the extremal
limit in (\ref{rad}), we get:
\begin{equation}
\frac{d^2\phi(r)}{dr^2} +
\frac{1}{r^2\Lambda} \frac{d(r^2\Lambda)}{dr}
{}~\frac{d\phi(r)}{dr} -
\frac{1}{\Lambda^2} \left[\frac{l(l+1)}{r^2} - E^2
\right]~\phi(r) ~=~0~~\label{exrad}.
\end{equation}
Expanding $\Lambda$ in powers of $GM/r$ and retaining terms to the
appropriate order,
this reduces to the Schwarzschild radial
equation (\ref{scrad}), and the scattering amplitude is once again
(\ref{amp}). Identical conclusions follow when one uses
the Einstein metric instead of the string metric.

\section{Scaling and Boundary Field Theory}
\label{sca}

So far, we have explicitly used the solutions of the dilaton
gravity action to model the point particles. In the second
section, the boosted particle was regarded as the source in the
background of which the slow particle scattered, while the latter
served as the source of a static spherically symmetric geometry
in section \ref{external}. In either case, the model failed except in the
extremal limit. Now, we approach the eikonal limit is a
`solution-independent way'. In other words, by imposing certain
kinematical restrictions, we suitably truncate the action of the
theory, such that it automatically incorporates the eikonal
kinematics. An important observation ensues to the effect that all
local degrees of freedom decouple from the theory, leaving behind
a residual boundary valued action. This has been demonstrated in
the case of general relativity and
electrodynamics separately in \cite{ver} and
\cite{jac} respectively. Our task would consist of two parts.
First, to show that in the Einstein-Maxwell framework,
the decoupling of the interactions take place
at the level of the action, as claimed in \cite{dm2} on the basis
of a heuristic analysis. Second, to investigate to what extent similar
arguments would hold for the case of the dilaton gravity action.
The advantage of this method is that
one does not have to resort
to explicit classical solutions at all.

We begin with the Einstein action
$$S_E~=~-\frac{1}{G}\int d^4x {\sqrt -g} R.$$
On choosing a gauge for the metric tensor such that
its longitudinal $(+,-)$ modes are manifestly decoupled from the
transverse modes $(i,j)$, and retaining only those configurations
which are consistent with the high momenta in the longitudinal
direction and low momenta in the transverse direction,
the Einstein action reduces to a action
on the boundary $\partial M$
of the two dimensional Minkowski subspace in the
following form \cite{ver}:
\begin{equation}
S_E \rightarrow S_{E[\partial M]}~=~\frac{1}{G}
\int {\sqrt g} \left( {\sqrt h}
R_h + \frac{1}{4} {\sqrt h} h^{ij} \partial_i g_{\alpha \beta}
\partial_j g_{\gamma \delta} \epsilon^{\alpha
\gamma}\epsilon^{\beta \delta}\right)~.
\label{ein}
\end{equation}
Here, all quantities pertaining to $g$ (with Greek indices)
and $h$ (Latin indices) are related to the
longitudinal and transverse subspaces respectively.
The metric components satisfy the constraints:
\begin{eqnarray}
h_{ij}~&=&~h_{ij}(x,y)~~,  \nonumber \\
g_{\alpha \beta}~&=&~\eta_{a b} \partial_{\alpha} X^a  \partial_\beta
X^b~~, \label{diffeo}
\end{eqnarray}
whereby $h_{ij}$ is no longer a propagating degree of freedom, and $g_{\alpha
\beta}$ is conformally flat upto diffeomorphisms of the longitudinal subspace.
Thus, only the boundary values of the diffeomorphism parameter $X^a$ remain as
the
surviving dynamical degrees of freedom in the eikonal limit.

The corresponding electromagnetic action in {\it flat}
space, namely
$$S_{EM}~=~-\frac{1}{4}\int d^4x F_{\mu \nu}F^{\mu \nu}$$
truncates (in the Lorentz gauge) to \cite{jac}:
\begin{equation}
S_{EM} \rightarrow S_{EM[\partial M]}~=~
\oint d\tau \int d^2 r_{\perp} \left(
\frac{1}{2} \Omega^{-}
\nabla^2 {\partial_\tau \Omega}^+
- \frac{1}{2} \Omega^{+}
\nabla^2 {\partial_\tau \Omega}^-
\right)~~,
\label{bdry}
\end{equation}
with the constraints for the fields:
\begin{equation}
F_{\pm} = 0~~;~~A_{\pm} = \partial_{\pm}\Omega~~;~~
\Omega (x)~=~\Omega^+ (x^+, {\vec r}_{\perp} ) + \Omega ^- (x^-,
{\vec r}_{\perp})~~.
\label{const}
\end{equation}
$A_i$ is a classical background and can be taken to be zero
without loss of generality. For both the gravity and
electromagnetic actions, it can be shown that the addition of the
terms representing interaction with matter currents does not alter
the topological nature of the action because the eikonal form of the
source currents can also be written as boundary terms. Incorporating
these terms, the $S$-matrix can be easily derived from the action
in the saddle
point approximation. The resulting scattering amplitude is
the expression (\ref{amp}) for gravity and $Gs \rightarrow -ee'$
for electromagnetism. In a short while we shall see how both these
terms can be incorporated in a single scattering amplitude
formula.
Finally, with the full Einstein-Maxwell action:
\begin{equation}
S~=~S_E + S_{EM}~=-~\int d^4x \sqrt{-g} \left[ \frac{R}{G} + \frac{1}{4}
g^{\mu\rho}g^{\nu\lambda}
F_{\mu \nu}F_{\rho
\lambda} \right], \label{emaction}
\end{equation}
the first (pure gravity) part once again reduces to the action on
the boundary. For the second (electromagnetism coupled to gravity)
part, the argument is more subtle. The results are best demonstrated in the
units of ref
\cite{ver}, where it was assumed that $dx^\mu~s$ were
dimensionless, whereas $g_{\mu \nu}$ had dimensions $L^2$, $L$ signifying a
length dimension.
For dimensional consistency, the other relevant quantities are
associated with the following dimensions:
$$\sqrt {-g} \sim L^4~,~g^{\mu \nu} \sim L^{-2}~;$$
$$d^4x \sim 1~,~x_{\mu} \sim L^2 $$
$$\partial_\mu \sim 1~,~
\partial^\mu \sim L^{-2}~;$$
$$A^{\mu} \sim L^{-2}~,~A_{\mu} \sim 1~~;$$
$$ F_{\mu
\nu}~ \sim 1 ~~~\&~~~F^{\mu \nu}~\sim L^{-4}~~.$$
Now let us consider the Maxwell action in an arbitrary space-time
background.
\begin{equation}
S_{EM}~=~-{1 \over 4} \int d^4x {\sqrt{-g}}F_{\mu \nu}  F^{\mu \nu}~.
\end{equation}
Splitting it up into the longitudinal, transverse and the mixed
parts, it takes the form:
\begin{equation}
S_{EM}~=~-{1 \over 4} \int d^4x{\sqrt {-g}}
\left[ F_{\alpha \beta}F^{\alpha \beta} +
2 F_{\alpha i}F^{\alpha i} + F_{ij}F^{ij}  \right]~.
\end{equation}
Now we scale the longitudinal components of all the tensors
by a small dimensionless
parameter $\lambda \sim \sqrt {t/s}$,
as $$x_{\alpha} \rightarrow
{\lambda}^2 x_{\alpha}~~;$$
$$F_{\mu \nu} \rightarrow F_{\mu \nu}~~,
{}~F^{\alpha \beta} \rightarrow {\lambda}^{-4}
F^{\alpha \beta}~~,~~F^{\alpha i} \rightarrow {\lambda}^{-2}F^{\alpha i}~~;$$
$$g_{\alpha \beta} \rightarrow
\lambda^2 g_{\alpha \beta}~~,~ \sqrt {-g} \rightarrow \lambda^2
\sqrt {-g}~~.$$
Note that the transverse components remain unchanged. The
rationale behind this scaling is that due to the high
center-of-mass energy $\sqrt {s}$, the longitudinal length scales
undergo a high Lorentz contraction which is incorporated in the
smallness of the corresponding scaled quantities. The
field components that survive after taking the limit $\lambda
\rightarrow 0$ in the action are to be regarded as the only relevant
degrees of freedom in the kinematical domain of interest.
With this in mind, the scaled electromagnetic action is:
\begin{equation}
S_{EM} \rightarrow -{1 \over 4}\int d^4x \sqrt {-g} {\lambda}^2 \left[ {1 \over
{\lambda}^4} F_{\alpha \beta}F^{\alpha \beta} + {1 \over {\lambda}^2}
2 F_{\alpha i}F^{\alpha i} + F_{ij}F^{ij} \right]~~.
\label{scale}
\end{equation}
As in the case of flat space-time, the first term is highly
oscillatory in the quantum partition function, which dictates the
dominant modes to be
$$F_{\pm} = 0~~,$$
admitting of the earlier solution
$$A_{\pm} = \partial_{\pm} \Omega~~.$$
As already mentioned, the transverse components of the gauge potential $A_i$
can be set
to zero since they decouple; the reduced action is thus
\begin{equation}
S_{EM}~=~ -{1 \over 2}\int d^4x \sqrt {-g}~
F_{\alpha i}F^{\alpha i}~~.
\label{scale1}
\end{equation}

Now, as pointed out after eq. (\ref{diffeo}), the metric $g_{\alpha \beta}$ is
conformally
flat in the longitudinal subspace, so that the conformally invariant quantity
$\sqrt{-g}
g^{\alpha \beta}$ can be transformed into the longitudinal Minkowski metric
$\eta^{\alpha
\beta}$ by local variations of $X^a$. Consequently, using eq. (\ref{diffeo}) we
can write
\begin{equation}
S_{EM}~=~-\frac{1}{2}\int d^2x_{\perp}\sqrt{h} h^{ij}~\int dx^+ dx^-~ F_{\alpha
i}F^{\alpha}_j~~. \label{scale2}
\end{equation}
On substituting the constraints (\ref{const}),
$$S_{EM}~=~\frac{1}{2} \int d^2x_{\perp}~\sqrt{h} h^{ij}~\int dx^+
dx^-\partial_i
\partial_{\alpha} \Omega \partial_j \partial ^{\alpha} \Omega~~.$$
As before, in the Lorentz gauge, this reduces to the action (\ref{bdry}) for
Minkowski
space scattering which enforces $h_{ij}=\delta_{ij}$.

In summary, the Einstein-Maxwell action in totality reduces to two
separate terms, representing the gravity and electromagnetic
interactions respectively,
\begin{equation}
S_E + S_{EM} \rightarrow S_{E[\partial M]} + S_{EM[\partial
M]}~~. \label{dec}
\end{equation}
Thus the $S$-matrix calculated from the total boundary action will
just be an incoherent superposition of the individual
$S$-matrices. This is the statement of decoupling that was
sought. For completeness, we give the expression for the
scattering amplitude of two point particles with charges $e$ and
$e'$ interacting via gravity and electromagnetism \cite{dm3} :
\begin{equation}
f(s,t)~=~{1 \over t}~\frac{\Gamma(1-iGs + iee')}
{\Gamma(1+iGs - iee')}\left(\frac{1}{-t}\right)^{-iGs + iee'}~,
\label{amp1}
\end{equation}
In effect, this means that we can replace the gravitational `coupling'
$Gs$ by the effective coupling constant $Gs - ee'$ in the presence of
electromagnetism.
It is remarkable that this decoupling is manifest already at the
level of the action, once the kinematical restrictions are imposed
on it.

We now move on to dilaton gravity. The
action that we must consider is (in the Einstein metric):
\begin{equation}
S_D~=~\int d^4x {\sqrt -g} \left[ -\frac{R}{G} + e^{-2\phi} F_{\mu \nu}F^{\mu
\nu} + 2 \partial_\mu \phi \partial^\mu \phi \right]~.
\label{straction}
\end{equation}
The first term is identical to the general relativity action
and independent of the dilaton field, yielding (\ref{ein}) once again. However,
the
interaction term involving the  Maxwell-Einstein-dilaton fields is no longer
amenable
to earlier simplifications. Although the scaling arguments will
still hold, the counterparts of Eqs. (\ref{scale1}) and
(\ref{scale2}) are respectively:
\begin{equation}
S_{EM}~=~ -{1 \over 2}\int d^4x \sqrt {-g}~
e^{-2\phi}F_{\alpha i}F^{\alpha i}~~
\label{scale3}
\end{equation}
and,
\begin{equation}
S_{EM}~=~-\frac{1}{2}\int d^2x_{\perp}\sqrt h h^{ij} \int
dx^+ dx^-  ~e^{-2\phi}F_{\alpha i}F^{\alpha}_j ~~
\label{scale4}
\end{equation}
The constraint $F_{\pm}=0$ will remain unchanged along with its
solution $A_{\pm}= \partial_{\pm}\Omega$. As before, $A_i$ is taken
to be zero. Thus the above equation becomes,
\begin{eqnarray}
S_{EM}&=&~ - \frac{1}{2} \int d^2x_{\perp} \sqrt h h^{ij} \int dx^+ dx^-
{}~ [ \partial_{\alpha} \left\{
e^{-2\phi} (\partial_i\Omega)(\partial_j\partial^{\alpha}\Omega)
\right\} - e^{-2\phi}(\partial_i\Omega) ( \partial_j
\partial_{\alpha}\partial^{\alpha} \Omega )  \nonumber \\
&+& e^{-2\phi}
(\partial_i\Omega) ( \partial_j \partial^{\alpha} \Omega)
\partial_{\alpha}\phi ]~~.
\end{eqnarray}
The first term is a total divergence and hence can be converted
into a boundary term. The second term can be made to vanish by
virtue of the Lorentz gauge condition. The new significant piece
is the last term, which is a `bulk' piece, dependent on the local
field coordinates. This term can neither be made to vanish, nor be
transferred to the boundary $\partial M$ for generic values of the
dilaton field. Thus, the local degrees of freedom fail to decouple
from the theory and eikonal approximation techniques used to calculate the
$S$-matrix can no longer be employed. These conclusions are of course not
dependent on the choice of coordinates. In terms of the string
metric, the dilaton couples to the scalar curvature as well as the
gauge fields. Thus, in this case, both the terms in the action would fail to
give pure boundary terms.

As in the previous sections, it is natural to investigate the
status of the above analysis in the extremal limit. However, here
since we are dealing with the action and not with the solutions,
it is not clear as to how one can implement the extremality
condition. Note however that the bulk term disappears for dilaton
configurations that are
independent of the null coordinates, i.e., when the dilaton ceases to be a
propagating
degree of freedom. As for example, consider the extremal limit of the black
hole solution.  The
solution for the dilaton field, derived from the action (\ref{straction}) is,
\begin{equation}
e^{2\phi}~=~e^{2\phi_0}\left(1-\frac{\alpha}{Mr}\right)~.
\label{dil}
\end{equation}
The extremality condition simplifies this to
\begin{equation}
e^{2\phi}~=~e^{2\phi_0}\left(1-\frac{2GM}{r}\right)~.
\end{equation}
Now, the eikonal limit requires that we take the particle masses
to be vanishingly small. Hence, on taking $M \rightarrow 0$ in the
above equation, we see that $\phi$ approaches its constant
asymptotic value {\it identically}. Thus the extremal dilaton solution
certainly is sufficient since the dilaton field is frozen at its extremal
value; but it
appears to be a bit of an overkill, since all one needs to eliminate the bulk
term is a
dilaton field depending only on the transverse coordinates.

\section{Resummation of ladder exchanges}
\label{pert}

Historically, the earliest approach to the eikonal approximation in
relativistic field
theory entailed analyses of an infinite
set of ladder-type exchange Feynman graphs in which the
momenta of the external lines are assumed to remain more or less fixed
on-shell, so that
virtual particles carried almost no momenta \cite{abar}. The motivation behind
this
restriction is the assumption that in the  high energy limit, there are well
defined
classical trajectories for the particles, which deviate only slightly from free
particle
trajectories. Ignoring standard radiative corrections, the infinite sum is seen
to admit
\cite{abar} of a closed
form expression, which indeed captures the leading behavior of the scattering
amplitudes
for high center-of-mass energies. A similar eikonal resummation for linearized
gravity,
involving ladder exchange of gravitons, was performed in
ref.\cite{kabat}, which reproduced the quantum mechanical result (\ref{amp}).
The
Feynman rules were derived form the following linearized gravity action:
\begin{eqnarray}
S_{LG}~&=&~\frac{1}{G}\int d^4x~ \frac{1}{8} h_{\mu \nu}\left[
\eta^{\mu \lambda}\eta^{\nu \sigma} + \eta^{\mu \sigma} \eta^{\nu
\lambda} - \eta^{\mu \nu} \eta^{\lambda \sigma} \right] \Box
h_{\lambda \sigma} \\  \nonumber
&+& \frac{1}{2} \chi \Box \chi + \frac{1}{2}
h_{\mu \nu} \left[ \partial^{\mu}\chi \partial^{\nu} \chi - \frac{1}{2}
\eta^{\mu \nu} \partial_{\sigma}\chi \partial^{\sigma} \chi
\right]~,
\end{eqnarray}
where the metric has been linearized as $g_{\mu \nu} = \eta_{\mu \nu} +
h_{\mu \nu}$. The scalar field $\chi$ corresponds to the particles
undergoing scattering. The eikonal amplitude obtained in this case, for
non-vanishing
masses, is given by \cite{kabat}
\begin{equation}
i {\cal M}(s,t)~\sim~{\sqrt{s(s-4m^2)} \over t} {\Gamma(1-i\alpha(s)) \over
\Gamma(i\alpha(s))}~, \label{geik} \end{equation}
where,
\begin{equation}
\alpha(s)~=~G {{(s-2m^2)^2 - 2m^4} \over \sqrt{s (s-4m^2)}}~~. \end{equation}
For $m=0$, this reduces to (\ref{amp}).

In the dilaton gravity case, if we start with the dilaton gravity action
coupled to the matter field $\chi$ in the string
metric,
\begin{equation}
S~=~\int d^4x \sqrt{-g}~e^{-2\phi} \left[ - \frac{R}{G} - 4
\partial_\mu \phi \partial^\mu \phi + F^2 - \frac{1}{2} \partial_\mu
\chi \partial^\mu \chi \right]~~,
\end{equation}
then the condition of the existence of the classical trajectory of
the test particles appears invalidated, since as already mentioned, for small
particle
masses,
the space-time singularity at $r=\alpha/M$ spreads indefinitely and traps any
other test
particle at arbitrarily large impact parameters. Thus a eikonal graph
calculation with
the above action is seemingly fraught with pitfalls. Despite these, we proceed
with linearizing
the dilaton field, as was done for the metric tensor. We write $\phi$ in the
form
$$\phi = \phi_0 + f~~,$$
where $f$ represents the small quantum fluctuations around the constant
asymptotic value $\phi_0$. Before embarking on perturbative
calculations with this simplified action, a
heuristic justification of this linearization may be given as follows. A rough
estimate
of the magnitude of $f$ can be made from the classical solution
(\ref{dil}),
$$f \approx \left|\phi - \phi_0 \right | \sim \left
|\ln\left(1-\frac{\alpha}{Mr}
\right)\right|~.$$
Demanding this to be small leads to the condition
$$\left|1-\frac{\alpha}{Mr}\right| \approx 1~~ \Longleftrightarrow
{}~~\left|\frac{\alpha}{Mr}\right| \approx 0~,$$
for arbitrary $r$. This of course means that $\alpha$ should
approach zero at least as $M^2$, which is the extremality condition. Hence a
linearized
approximation seems reasonable in the extremal limit.

To leading orders in the graviton and dilaton fluctuations, the dilaton gravity
action now
becomes
\begin{eqnarray}
S~&=&~  \frac{e^{-2\phi_0}}{G}  \int d^4 x
(1- 2 f) \frac{1}{8} h_{\mu \nu} \left[ \eta^{\mu \lambda} \eta^{\nu
\sigma} + \eta^{\mu \sigma} \eta^{\nu \lambda} - \eta^{\mu \nu}
\eta^{\lambda \sigma} \right] \Box h_{ \lambda \sigma} \nonumber
\\
 &-&~  e^{-2\phi_0} \int d^4x \left( 1 + \frac{1}{2}
h_{\alpha}^{~\alpha}\right) \left(1 - 2 f\right ) \left[ - 4 \partial_\mu f
\partial^\mu f + F^2 + \partial_\mu \chi \partial^\mu \chi
\right]~~.
\end{eqnarray}
Since the graviton and photon ladder summations are known, we concentrate on
the
dilaton-matter field interactions, given by the last term . The new
momentum dependent
$(\chi-\chi-f)$ vertex is associated with the factor $-2p\cdot p'$,
where $p$ and $p'$ are the momenta associated with the two
$\chi$ lines. They give rise to an
infinite set of ladders with intermediate dilaton exchanges.
Since these can be summed in a fairly straightforward manner, we
simply give a schematic derivation of the final result. The Born
amplitude (corresponding to a single dilaton exchange) is
\begin{equation}
i {\cal M}_{Born}~=~ \frac{i p_1^2 p_2^2}{(p_1 - p_3)^2 -i
\epsilon}~.
\end{equation}
Here, $p_1$ and $p_2$ are the incoming and
$p_3$ and $p_4$ are the outgoing
4-momenta. They are related by the constraint $p_1 + p_2 - p_3 -p_4
=0$.
For the next higher order ladder,
there are four distinct diagrams depending on the momentum
labels for the two exchanged particles. Using the eikonal form of
the external matter propagators \cite{abar,kabat}, namely
$$
\frac{1}{(p+k)^2+m^2-i\epsilon} \approx \frac{1}{2p\cdot k -
i\epsilon}~,$$
the one loop
amplitude is,
$$
p_1^4 p_2^4 \int {d^4 k \over (2 \pi)^4}~\frac{1}{k^2 - i
\epsilon}~\frac{1}{(p_1 - p_3 -k )^2 - i \epsilon} \nonumber \\
$$
$$
\times \frac{1}{2}
[ \frac{1}{-2p_1 \cdot k - i \epsilon}~\frac{1}{ 2 p_2 \cdot
k - i \epsilon}  + \frac{1}{ -2p_1 \cdot k - i \epsilon}~\frac{1}{
- 2p_4 \cdot k - i \epsilon}
$$
$$
+ \frac{1} { 2p_3 \cdot k - i
\epsilon}~\frac{1}{ 2 p_2 \cdot k - i \epsilon} + \frac{1}{ 2 p_3
\cdot k - i \epsilon}~\frac{1}{ - 2p_4 \cdot k - i \epsilon}
]
$$
By doing the combinatorics carefully, it can be shown that the
infinite set of ladders exponentiate to give the final amplitude
as
\begin{equation}
i {\cal M}~=~-p_1^2p_2^2 \int d^4x e^{-(p_1-p_3) \cdot x}~\Delta
(x)~\frac{e^{i\psi}-1}{\psi}~,
\label{born}
\end{equation}
where $\Delta(x)$ is the fourier transform of the dilaton
propagator and
$$
\psi ~=~ -p_1^2 p_2^2 \int {d^4 k \over (2 \pi)^4}~e^{ik\cdot x}
{}~\frac{1}{k^2 - i
\epsilon} ~~~~~~~~~~~~~~~~~~~~~~~~~~~~~~~
$$
$$
{}~~~~~~~\times~
[~\frac{1}{-2p_1 \cdot k - i \epsilon}~\frac{1}{ 2 p_2 \cdot
k - i \epsilon}  + \frac{1}{ -2p_1 \cdot k - i \epsilon}~\frac{1}{
- 2p_4 \cdot k - i \epsilon}
$$
$$
+ \frac{1} { 2p_3 \cdot k - i
\epsilon}~\frac{1}{ 2 p_2 \cdot k - i \epsilon} + \frac{1}{ 2 p_3
\cdot k - i \epsilon}~\frac{1}{ - 2p_4 \cdot k - i \epsilon}
{}~]~.
$$
Assuming small momentum transfers, we can take $p_1 \approx p_3$
and $p_2 \approx p_4$, to obtain
$$\psi~=~-\frac{p_1^2 p_2^2}{16 \pi E p } \ln \mu x_\perp~.$$
Here $x_{\perp}$ is the transverse coordinate, $(E,\pm p)$ are the
four-momentum vectors of the two particles in the center-of-mass
frame and $\mu$ is an irrelevant mass parameter. With this, the
explicit evaluation of $\cal M$ in (\ref{born}) leads to
\begin{equation}
i {\cal M}~=~\frac{i p_1^2 p_2^2}{-t}~
\frac{\Gamma \left( 1 -
i p_1^2 p_2^2 / 32 \pi Ep \right)}
{\Gamma \left( 1 +
i p_1^2 p_2^2 / 32 \pi Ep \right)}~\left(\frac{4 \mu^2}{-t}
\right)^{-i \frac{p_1^2 p_2^2}{32 \pi Ep}}~,
\end{equation}
where $-t$ is the square of the momentum transfer. Now, plugging
in the on shell conditions $p_1^2,p_2^2 = m^2$, the above
amplitude decays to zero for vanishing particle masses.
This means that these ladders do not contribute to the scattering
amplitude at all ! Thus we are left with the original set of
matter-graviton and matter-photon ladder-diagrams of refs.\cite{kabat,jac}
and the corresponding finite scattering
amplitude for Einstein-Maxwell theory (\ref{amp1}).

It now seems that the
pathologies that we had encountered earlier have
disappeared. Note however that the preceding results would only
hold when the dilaton fluctuations are small enough for linearization to go
through,
i.e., $|\phi-\phi_0|~\ll~1$ (in Planck units). Now, in the low energy limit
of string theory, the string coupling parameter $g_s$ is usually
related to the asymptotic value of the dilaton, $g_s \equiv \exp{
\phi_0}$. In the regime of perturbative string theory one must have
$g_s~ \ll 1$, which implies
that $\phi_0$ itself should be large and negative (in Planck units), i.e.,
$|\phi_0|~\gg~1$. It is not clear that these dual requirements are compatible.
Thus, our
linearization of the dilaton gravity action may not correspond to the
perturbative domain of string theory. But if we now relax this
restriction to include large $g_s$ regimes, then the
the linearization is perfectly justified and there is no problem with
resummation of
dilatonic ladder exchanges. Since certain extremal black hole solutions of
string theory
\cite{sen} have been advertized as exact quantum states not subject to the
perturbative
restriction
$g_s~\ll~1$, it is perhaps not surprising that Planckian scattering of point
particles,
which is inherently non-perturbative in nature, is reasonable only outside the
perturbative
regime of string theory.

\section{Conclusion}
\label{concl}

We begin this section with a survey of our principal findings. The curvature
singularity
away from the origin in the non-extremal charged dilaton black hole metric is
shown to
be responsible for the absence of a plane-fronted gravitational shock wave,
when such a
black hole is Lorentz-boosted to luminal velocities. Instead of a single plane
($x^-=0$ in the Schwarzschild case), the singular geometry in the Planckian
eikonal
limit consists of a three dimensional region whose thickness is proportional to
the
dilatonic charge $\alpha \equiv Q^2 \exp{(-\phi_0)}$. Consequently, Planckian
scattering
amplitudes in this model can no longer be computed using the simple techniques
of ref.
\cite{thf}. The problem resurfaces in the external metric approach in that the
radial component of the particle equation of motion does not reduce to a
Schr\"odinger-like
equation in the eikonal approximation. In fact, the discontinuities in the
coefficients
of this equation in the relevant kinematical limit render the equation
unsolvable.
Remarkably, in both approaches, the malady disappears upon imposing the
extremal limit;
in the first (heuristic) approach, the dilaton charge simply shrinks to zero
upon
boosting, thereby yielding the same plane-fronted gravitational shock wave as
in the
Schwarzschild case. An identical situation ensues in the external metric
formalism,
where the discontinuities previously preventing the solution of the quantum
equation of
motion are now gone. Since the static extremal dilatonic black hole metric
looks quite
different from the Schwarzschild metric, the end-result is a pleasant surprise.

The alternative approach involving identification of the degrees of freedom
participating in eikonal scattering and an effective field theory of these
degrees of
freedom a la\'~ref. \cite{ver}, has also been pursued for the dilaton gravity
action.
Indeed, unlike in the case of the Einstein-Hilbert and Maxwell actions, this
action does
{\it not} reduce in the appropriate scaling limit to a `boundary' field theory.
The
offending terms disappear for non-propagating dilaton configurations such as
would
appear for extremal black hole solutions in the massless limit. The situation
is however
quite different for the standard field theoretic approach to the eikonal of
summing
ladder exchange Feynman graphs. In this case, a linearized approximation to the
dilaton
gravity action, retaining terms only upto quadratic in the dilaton field, does
indeed
yield a summed amplitude of ladders and crossed ladders in a closed form in the
eikonal
kinematical domain. The problem shows up in a rather subtle manner: the
restriction on
the asymptotic value of the dilaton field from string perturbation theory is
not
compatible with the requirement of small dilaton fluctuations around the
asymptotic value
necessary for linearization of the action (and the subsequent derivation of the
eikonal
amplitude).

The above analyses point unambiguously to the fact that extremal
black holes play a very special role in eikonal scattering. Recall that
our motivation to consider dilaton gravity was to model charged
point particles as sources of the dilaton gravity metric instead
of the canonical Reissner-Nordstr\"om  metric. The reason was of course
that the low energy string equations of motion naturally give rise
to the former. However, this modelling seems to work only in
the extremal limit. Perhaps this is the manner that string theory, which gives
rise
inexorably to dilaton gravity at low energies as an effective theory of
gravitation, also
cures the problems that go with it. The central role played by extremal black
holes is
emphasized time and again in recent literature on duality, because of the
strong
possibility of their being elementary string excitations
\cite{sen}. Our work stresses this further
in terms of non-perturbative behavior in the eikonal limit.

\end{document}